\documentclass{iau}

\usepackage{amsmath}
\usepackage{graphicx}
\usepackage{multirow}

\begin{document}

\lefttitle{Kacharov \& Cioni}
\righttitle{IAU Symposium 379}

\jnlPage{1}{7}
\jnlDoiYr{2023}
\doival{10.1017/xxxxx}

\aopheadtitle{Proceedings of IAU Symposium 379}
\editors{P. Bonifacio, M.-R. Cioni, F. Hammer, M. Pawlowski, and S. Taibi, eds.}

\title{Equilibrium dynamical models for the Large Magellanic Cloud}

\author{N. Kacharov$^1$ \& M.-R. L. Cioni$^1$}
\affiliation{$^1$ Leibniz Institute for Astrophysics, An der Sternwarte 16, 14482 Potsdam, Germany}

\begin{abstract}
The Large Magellanic Cloud (LMC) has a complex dynamics driven by both internal and external processes.
The external forces are due to tidal interactions with the Small Magellanic Cloud and the Milky Way, while internally its dynamics mainly depends on the stellar, gas, and dark matter mass distributions.
Despite the overall complexity of the system, very often simple physical models can give us important insights about the main driving factors.
Here we focus on the internal forces and attempt to model the proper motions of $\sim10^6$ stars in the LMC as measured by Gaia Data Release 3 with an axisymmetric dynamical model, based on the Jeans equations.
We test both cored and cusped spherical Navarro-Frenk-White dark matter halos to fit the LMC gravitational potential.
We find that this simple model is very successful at selecting a clean sample of genuine LMC member stars and predicts the geometry and orientation of the LMC with respect to the observer within the constraint of axisymmetry.
Our Jeans dynamical models describe well the rotation profile and the velocity dispersion of the LMC stellar disc, however they fail to describe the motions of the LMC bar, which is a non-axisymmetric feature dominating the central region.
We plan a triaxial Schwarzschild approach as a next step for the dynamical modelling of the LMC.
\end{abstract}

\begin{keywords}
Stellar dynamics, Jeans models, LMC
\end{keywords}

\maketitle

\section{Introduction}

The Large Magellanic Cloud (LMC) is the largest satellite galaxy of the Milky Way, however its total mass has been a subject of debate.
With a stellar mass $\sim2.7\times10^9\,\rm M_{\odot}$ and gas mass $\sim0.5\times10^9\,\rm M_{\odot}$ \citep[][total luminosity of $\sim1.3\times10^9\,\rm L_{\odot}$]{vandermarel+2002}, it is clear that the LMC's total mass is dominated by its dark matter (DM) halo.
Multiple lines of evidence \citep[from internal kinematics, stellar streams, number of LMC satellites, effects from interactions with the Small Magellanic Cloud (SMC) and the Milky Way; see the review by][and references therein for more details]{vasiliev2023} suggest that the total mass of the LMC is in the range of $1-2\times10^{11}\,\rm M_{\odot}$.
This implies a mass-to-light ratio in the order of $100\,\rm\frac{M_{\odot}}{L_{\odot}}$, meaning that the LMC DM halo has not been stripped through the interaction with the Milky Way.

While it is indeed close to impossible to use kinematic tracers to study the dynamics of the outer regions of the LMC due to them being highly perturbed from its multiple interactions and close encounters with the Milky Way and the SMC \citep[e.g.][]{choi+2022}, the relatively high mass density, preserved until the present day, makes equilibrium dynamical models a viable option to understand the galaxy's dynamics in its inner regions.
Furthermore, any structure and details in the residuals of the equilibrium models can prove to be a valuable insight into the nature and significance of the non-equilibrium effects in the internal structure of the LMC.

We envision that a good handle of the dynamics in the innermost regions of the LMC through high quality data and careful modelling could facilitate the search for a possible massive black hole in the centre of the galaxy.

The LMC has a prominent bar structure that dominates the surface brightness and stellar kinematics in the inner region.
The bar appears off-centred with respect to the centre of rotation of the H\,I gas, however the centre of rotation of the stellar component seems to be much better aligned with the photometric centre, close to the centre of the bar \citep{luri+2021}.
The bar likely has a triaxial shape and is a source of non-axisymmetric perturbations in the centre of the galaxy, that may drive gas flows and trigger star formation.
It was most likely formed from the mutual interactions between the LMC and the SMC \citep{besla+2012}.
For the proper modelling of this structure we need to move away from the assumption of axisymmetry and use alternative methods like the Schwarzschild orbit superposition approach \citep{vandenbosch+2008, tahmasebzadeh+2022}.  

Here we present only preliminary equilibrium models of the LMC, based on the Jeans equations as a first step towards our goal of a complete dynamical model of the inner regions of the galaxy, using state of the art kinematic data sets.
A similar analysis, based on Gaia Data Release 2 data was published by \citet{vasiliev2018}.

\section{Data and construction of the Jeans models}

In this study we use Gaia Data Release 3 (DR3) proper motion (PM) data in the footprint of the VMC survey of the LMC \citep{cioni+2011}.
We consider only stars that are detected in both VMC and Gaia, and have valid Gaia PM measurements.
We additionally require that the Gaia DR3 PM uncertainties in the RA and DEC directions are less than $0.1$\,mas\,yr$^{-1}$.
This leaves us with $\sim10^6$ stars down to $\rm G = 18$\,mag to work with.

We assume a distance modulus to the LMC $\rm (m-M) = 18.52$\,mag \citep{crandall+ratra2015}, which corresponds to a distance of $50.58$\,kpc.
We fix the centre of the LMC at $\rm \alpha_0=81.07$\,deg and $\rm \delta_0=-69.41$\,deg from \citet{luri+2021} and convert the stellar coordinates and PMs to an orthographic projection $\rm (\alpha, \delta, \mu_{\alpha}. \mu_{\delta}) \rightarrow (X, Y, \mu_X. \mu_Y)$.
We also correct the PMs for perspective effects, as in \citet{vandeven+2006}.

For the surface brightness distribution of the LMC, we assume an exponential profile with $\rm r_e = 98.7$\,arcmin \citep{gallart+2004} and a total luminosity of $1.31\times10^9\,\rm L_{\odot}$ from the NASA/IPAC Extragalactic Database (NED), which we represent with a Multi-Gaussian Expansion (MGE).

With the Jeans dynamical approach we can model the LMC as an axisymmetric flattened ellipsoid.
The Jeans equations link the unknown gravitational potential to the stellar density and motions, which are derived from observations.
In this study we make use of the Jeans Anisotropic Modelling (JAM) code by \citet{cappellari2008}, assuming a cylindrical alignment of the velocity ellipsoid.

We model the gravitational potential with a generalised Navarro-Frenk-White (NFW) DM profile, where we can change the centre slope ($\gamma$).
This profile is characterised by two parameters (scale radius - $r_0$ and central density - $\rho_0$), which are free parameters in our model.
At first we consider a cuspy DM distribution by fixing $\gamma=1$.

The rest of the free parameters in our models are related to the geometry (inclination $i$, rotation axis orientation $\theta$, flattening $q = \frac{b}{a}$) and kinematical properties (anisotropy $\beta_z$ and level of rotation $\kappa$) of the LMC.
We also leave the bulk PMs in the $x$ and $y$ directions as free parameters, and one last free parameter $\epsilon$ controls the fraction of non-member stars.

We use a maximum likelihood approach on the discrete PM measurements and the Markov Chain Monte Carlo method to fit the model to the data.
The total likelihood per star is given by the joint probability of it following the adopted exponential surface density distribution (spatial likelihood) and having kinematical properties consistent with the predictions of the Jeans equations (dynamical likelihood).

\section{Results}

\begin{figure}
    \includegraphics[scale=.25]{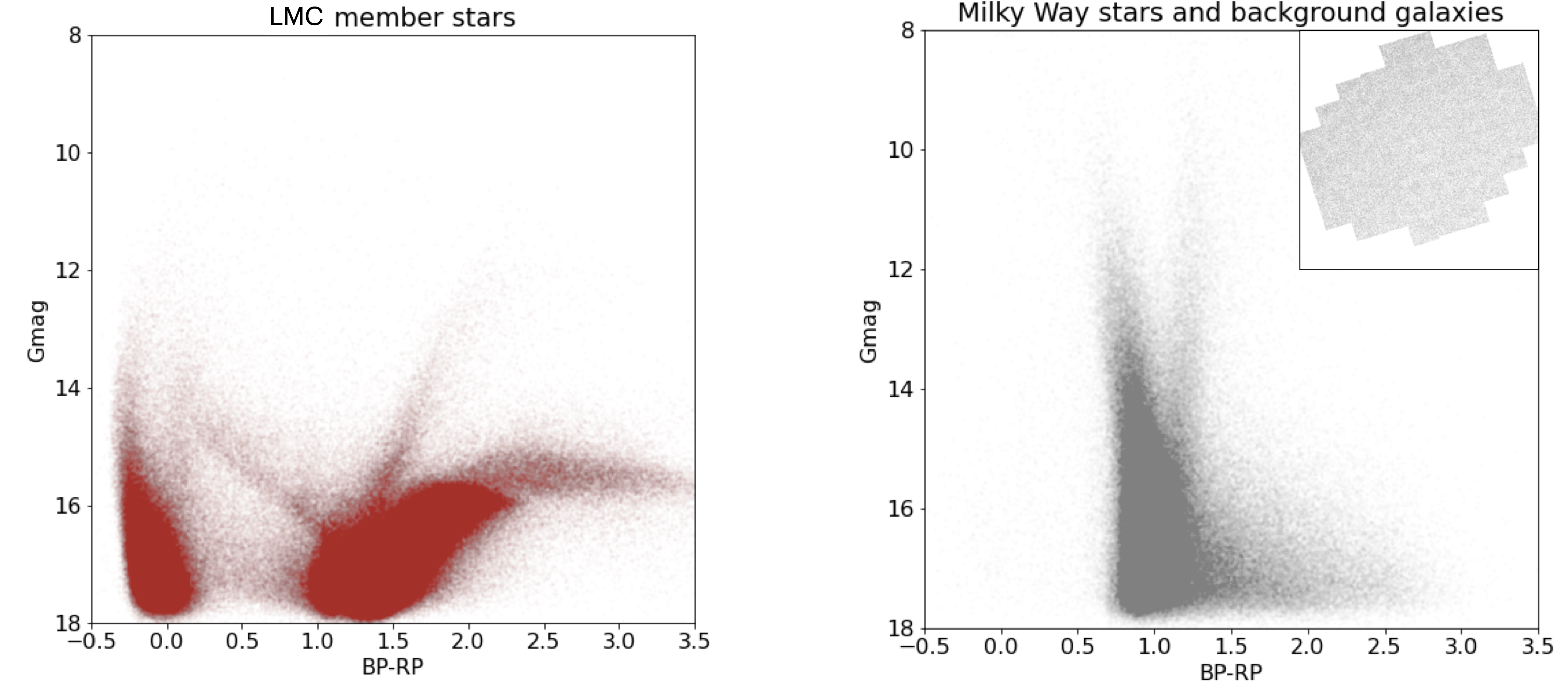}
    \caption{Left: a CMD of stars with high probability to be genuine LMC members. Right: a CMD of targets with high probability to be foreground Milky Way contaminants or background galaxies. The inlet shows their flat spatial distribution.}
    \label{fig:mem}
    \end{figure}

A first outcome of our converged model fit is that we can easily compute the probability of each target to belong to either the LMC or the foreground / background population.
Figure \ref{fig:mem} shows the colour-magnitude diagram (CMD) of stars with high probability of being genuine LMC members on the left and entries with high probability of belonging to the foreground / background on the right, which also have a very flat spatial distribution throughout the VMC footprint, as expected.
The left hand side CMD clearly shows the dense regions occupied by LMC red giant branch (RGB) and main sequence (MS) stars, but also the less dense sequences of the asymptotic giant branch (AGB), supergiants, red and blue He burning phases are easily identifiable.

We find a best fit inclination angle $i = 33.2$\,deg in good agreement with the kinematical analysis from the \citet{luri+2021} and an intrinsic flattening of $\rm q_{intr} = 0.2$.
Our best fit estimates for the LMC mean PMs are $\rm \mu_{\alpha}^0\cos(\delta) = 1.87\,mas\,yr^{-1}$ and $\rm \mu_{\delta}^0 = 0.36\,mas\,yr^{-1}$.
The statistical uncertainties of these parameters are very small, so they are fully dominated by systematic errors, such as choice of central coordinates, distance, surface density distribution, shape of the potential, etc.
For instance changing the inner slope ($\gamma$) of the DM density profile is tightly correlated with our estimates of the LMC intrinsic flattening and inclination angle.
In a dynamical model, where $\gamma = 0$ (DM core), we get a much thinner LMC disc with $\rm q_{intr} = 0.05$ and $\rm i = 32.65$\,deg.
On the other hand if we let $\gamma$ to be a free parameter, we find best-fit $\gamma=1.5$ and $\rm q_{intr} = 0.3$.

\begin{figure}
    \includegraphics[scale=.22]{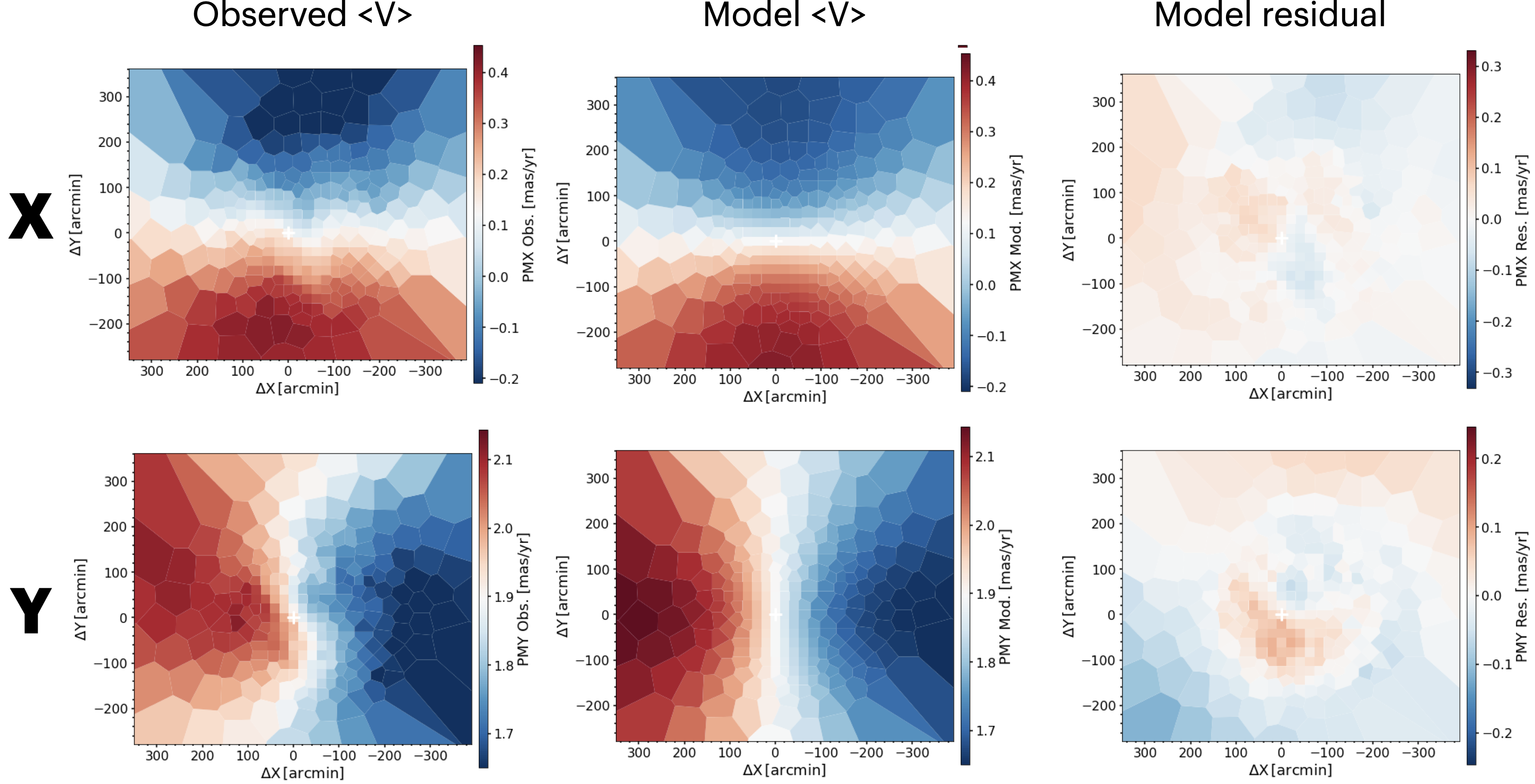}
    \caption{Observed and modelled rotation profile of the LMC. The top row shows the observed, modelled and residual PM in the direction of the horizontal axis ($\rm\mu_X$). Similarly, the bottom row shows the observed, modelled, and residual PM along the vertical axis ($\rm\mu_Y$).}
    \label{fig:vel}
    \end{figure}

\begin{figure}
    \includegraphics[scale=.22]{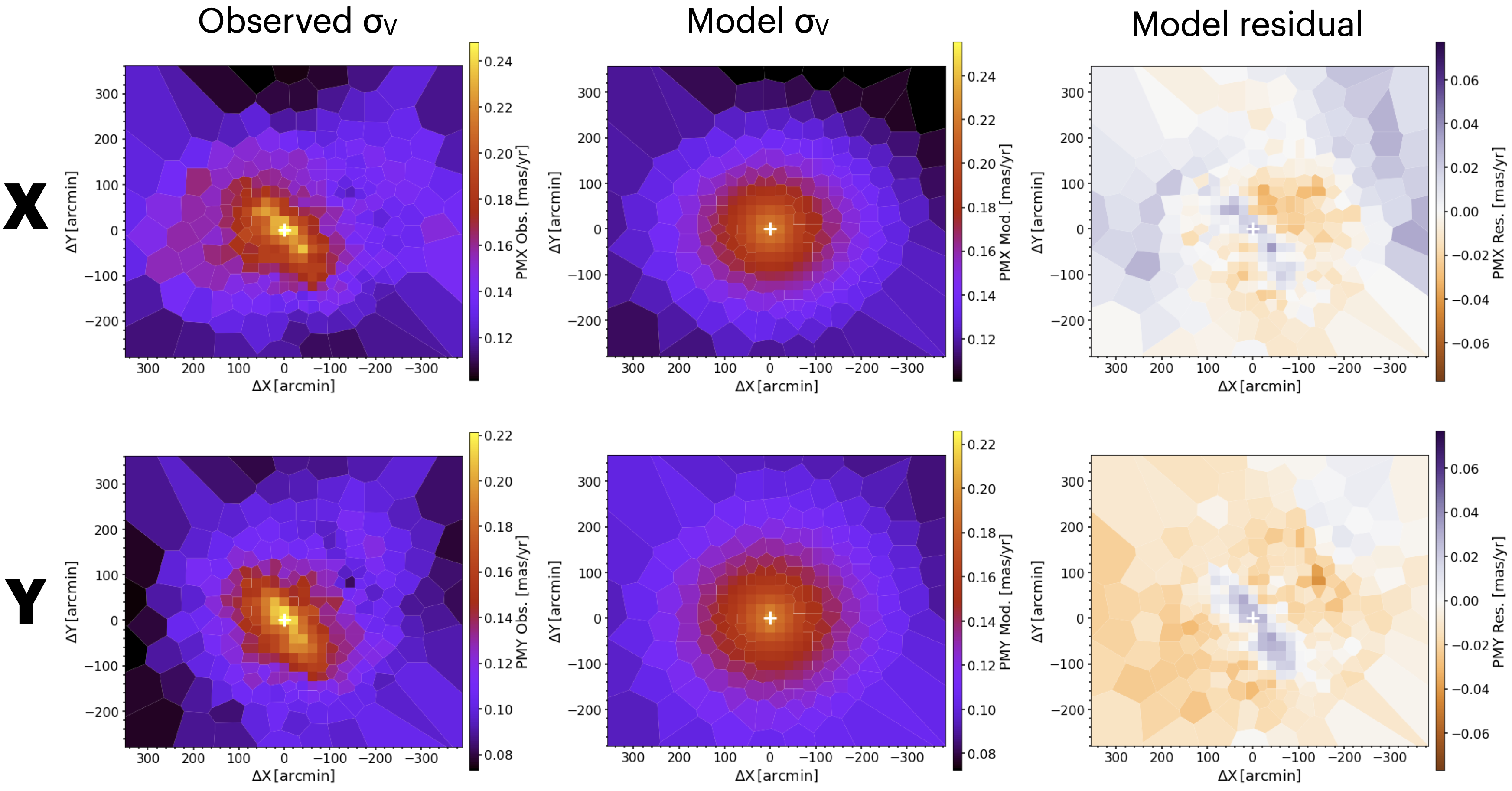}
    \caption{Observed and modelled velocity dispersion profile of the LMC. The top row shows the observed, modelled and residual PM velocity dispersion in the direction of the horizontal axis ($\rm\mu_X$). Similarly, the bottom row shows the observed, modelled, and residual velocity dispersion PM along the vertical axis ($\rm\mu_Y$).}
    \label{fig:sigma}
    \end{figure}

Figures \ref{fig:vel} and \ref{fig:sigma} show how well our axisymmetric Jeans dynamical model fits the observed kinematics of the LMC in PM space - the galaxy's rotation and velocity dispersion profiles, respectively.
The field of view in the figures is rotated such that the symmetry axis is along the vertical $y$ direction.
The rotation profile is fit generally quite well with median residuals in the order of only $1-2\%$.
While the Jeans model also fits well the dispersion profile in the outer region, dominated by the LMC disc, the inner region shows strong non-axisymmetric features, which cannot be reproduced with Jeans models.
This is because the kinematics in the central parts of the galaxy is dominated by the LMC bar.
Besides not fitting well the inner region velocity dispersion spatially, the Jeans model also underestimates the central velocity dispersion.
It was noted that the observed high-velocity dispersion in the centre of the LMC is artificially biased due to crowding problems with the Gaia DR3 data \citep{libralato+2023}, which could be fixed by using PM measurements from the VMC survey, which has higher spatial resolution.
The non-axisymmetric effects of the bar are also visible in the rotation map, where they show as apparent wiggles of the rotation axis.
Overall, we conclude that the axisymmetric Jeans model is adequate at describing the LMC disc, but it is not appropriate for the inner region of the galaxy, dominated by a non-axisymmetric bar feature.

\begin{figure}
    \includegraphics[scale=.28]{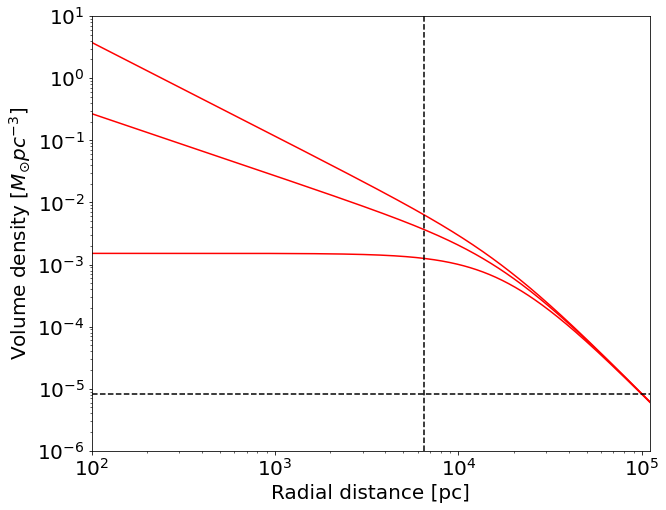}
    \includegraphics[scale=.28]{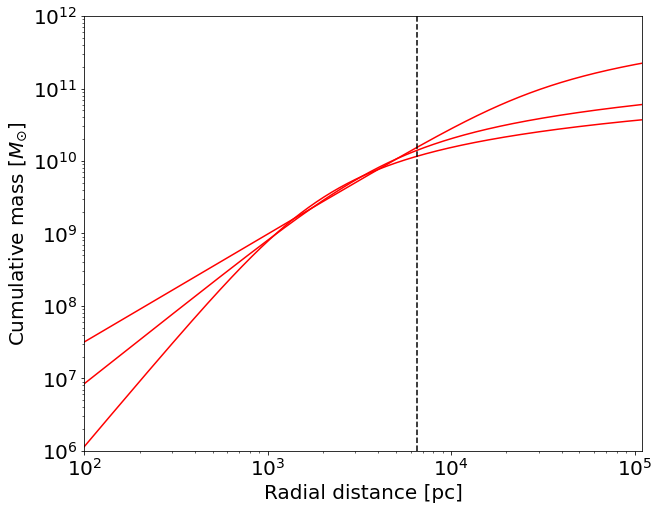}
    \caption{Left: Best fit NFW radial density profiles with central density slopes $\gamma=0$, $\gamma=1$, and best-fit $\gamma=1.5$. Right: Best fit cumulative mass profiles for the three NFW density profiles with varying central density slopes. The horizontal line is at $\rm 200\rho_{crit}$ level and the vertical line shows the extend of our data.}
    \label{fig:mass}
    \end{figure}

Finally, we turn to the mass estimates for the LMC from our Jeans models.
Our potential tracer data extends out to $7$\,kpc from the centre of the galaxy.
The enclosed DM mass within this radius is $\sim10^{10}\,\rm M_{\odot}$ for all dynamical models, independent of the central density slope ($\gamma$, see Figure \ref{fig:mass}).

We can also calculate $\rm M_{200}$ from the best-fit central density, scale radius, and central density slope of the NFW DM profiles.
Our default model with $\gamma=1$ gives a $\rm M_{200}$ estimate of $3.3\times10^{10}\,\rm M_{\odot}$ and the cored model ($\gamma=0$) gives $\rm M_{200} = 2.3\times10^{10}\,\rm M_{\odot}$.
These mass estimates are significantly lower than independent measurements, that put the total mass of the LMC closer to $1.5\times10^{11}\,\rm M_{\odot}$ \citep[see Figure 1. in][]{vasiliev2023}.
However, we mentioned above that our Jeans models generally underestimate the central velocity dispersion.
This biases the best-fit NFW profile, which needs to be extrapolated to the outskirts of the galaxy and hence we obtain a lower total mass estimate.
When we let $\gamma$ to be a free parameter in the fit, we find best-fit $\gamma=1.5$, which matches better the observed velocity dispersion in the inner region and we get $1.1\times10^{11}\,\rm M_{\odot}$ in much better agreement with independent estimates.

In order to get a good handle of the total mass of the LMC we need to be able to model the stellar tracers at a wide range of radial distances, which also includes the LMC bar - a triaxial feature, that cannot be modelled with the Jeans approach.

\section{Conclusion and outlook}

While the axisymmetric Jeans models cannot provide a good physical representation of the dynamical state of the LMC as a whole, they could prove useful in describing its disc.
The Jeans models are successful in recovering the orientation and geometry of the LMC disc under the axisymmetry assumption, fit well its kinematics in PM space, and distinguish probabilistically genuine member stars from foreground Milky Way contaminants and background galaxies.

For the inner region of the LMC, however, which is dominated by the bar structure, we need to resort to triaxial models.
It is therefore interesting to explore the Schwarzschild orbit superposition approach in dynamical modelling with {\sc dynamite} \citep{vandenbosch+2008, thater+2022}.
Our goal is to obtain an orbit-based model of the LMC, where we fit the Gaia proper motions and surface density distributions of the two main components of the galaxy - an axisymmetric disc and a triaxial bar, simultaneously.
This type of setup of Schwarzschild models were recently pioneered by \citet{tahmasebzadeh+2022}.

It should be noted, however, that the orbit superposition method still works under the constraints of a dynamical equilibrium.
The LMC is plunging into its deepest peri-centre point into the Milky Way, where maximal tidal shocks and ram pressure is expected.
In that context the LMC is quite unique in having a smaller gas disc than a stellar disc.
Moreover, the galaxy has recently been in collision with the SMC $\sim250$\,Myr ago \citep{choi+2022}.
It would be interesting to explore the possibility to introduce perturbations in the dynamical models to study the deviations from equilibrium.
However, a good equilibrium model could also be subtracted from the actual observations and the remaining residuals may give us valuable information about the magnitude of the complex dynamical forces that pull the LMC out of equilibrium.

Furthermore, the off-center LMC bar is likely a transient structure due to the interaction with the SMC \citep[see e.g. the simulations in][]{besla+2012}.
So it is likely that any equilibrium modelling of the bar can only capture part of its full complexity.

\section*{Acknowledgements}
We thank Francois Hammer and Roeland van der Marel for the insightful discussion after the presentation.
The answers to their questions and comments are incorporated in the proceeding.

\section*{Discussion after the presentation}
{\it Roeland van der Marel:} Congratulations on your nice work adding much needed realism to simple circular orbit models. Two comments:

(1) The velocity dispersions reported near the centre of the LMC from Gaia eDR3 by \citet{luri+2021} are likely very significant overestimates.
See e.g. the recent proper motion data from HST and JWST and discussion in \citet{libralato+2023}.
So one would not expect a realistic model to reproduce them.

{\it Nikolay Kacharov:} We could cross-match the Gaia eDR3 PMs with PMs from the VMC survey to complement the data.

(2) The off-centre LMC bar is likely a transient structure due to the interaction with the SMC.
See e.g. the simulations in \citet{besla+2012}.
So any equilibrium modelling of the bar can likely only capture part of its full complexity.

{\it Francois Hammer:} The LMC is plunging at its deepest peri-center point into the Milky Way, where maximal tidal shocks and ram pressure is expected, and indeed the one LMC peculiarity is to be quite unique in having a smaller gas disk than stellar disk.
Moreover, its main stellar axis is off by $\sim50$\,deg from its kinematics, and it has recently been in collision with the SMC $\sim250$\,Myr ago.
Would you think possible to enter some effect of this disequilibrium to verify how it can change your modelling?

{\it Nikolay Kacharov:} The orbit-superposition technique works under the assumption of dynamical equilibrium, however, examining the model residuals can inform us about the magnitude of deviation from this state.

\end{document}